# In vivo imaging and tracking of individual nanodiamonds in drosophila melanogaster embryos

David A. Simpson[1,2*], Amelia J. Thompson[3], Mark Kowarsky[1], Nida F. Zeeshan[3], Michael S. J. Barson[1], Liam T. Hall[1,2], Yan Yan[4], Stefan Kaufmann[1], Brett C. Johnson[1,5], Takeshi Ohshima[5], Frank Caruso[4], Robert E. Scholten[1,6], Robert B. Saint[3], Michael J. Murray[3], Lloyd C. L. Hollenberg[1,2]

[1]*School of Physics, The University of Melbourne, Victoria 3010, Australia.*
[2]*Centre for Quantum Computation and Communication Technology, School of Physics, The University of Melbourne, Victoria 3010, Australia.*
[3]*Department of Genetics, The University of Melbourne, Victoria 3010, Australia.*
[4]*Department of Chemical and Bio-molecular Engineering, University of Melbourne, Victoria 3010, Australia.* [5]*Radiation Effects Group, Japan Atomic Energy Agency, Takasaki, Gunma 370-1292, Japan.*
[6]*Centre for Coherent X-ray Science, School of Physics, University of Melbourne, Victoria 3010, Australia.*

[*]*simd@unimelb.edu.au*

**Abstract:** In this work, we incorporate and image individual fluorescent nanodiamonds in the powerful genetic model system Drosophila melanogaster. Fluorescence correlation spectroscopy and wide-field imaging techniques are applied to individual fluorescent nanodiamonds in blastoderm cells during stage 5 of development, up to a depth of 40 µm. The majority of nanodiamonds in the blastoderm cells during cellularisation exhibit free diffusion with an average diffusion coefficient of $(6 \pm 3) \times 10^{-3}$ µm$^2$/s, (mean ± SD). Driven motion in the blastoderm cells was also observed with an average velocity of $0.13 \pm 0.10$ µm/s (mean ± SD) µm/s and an average applied force of $0.07 \pm 0.05$ pN (mean ± SD). Nanodiamonds in the periplasm between the nuclei and yolk were also found to undergo free diffusion with a significantly larger diffusion coefficient of $(63 \pm 35) \times 10^{-3}$ µm$^2$/s (mean ± SD). Driven motion in this region exhibited similar average velocities and applied forces compared to the blastoderm cells indicating the transport dynamics in the two cytoplasmic regions are analogous.

## 1. Introduction

Tracking the motion of individual particles in living systems has over the past two decades, provided enormous insight into the bio-molecular activity occurring at the intra-cellular level. Single particle tracking (SPT) techniques have been at the forefront of this revolution, allowing exploration of single motor protein stepping [1-3], DNA polymerisation [4] and cell membrane motility [5, 6]. The principle behind single particle tracking involves monitoring and tracking the position of an individual fluorescent probe over time with high spatial resolution. The effectiveness of the technique is often limited by the properties of the fluorescence probe itself. Traditional fluorescent probes such as single molecules, fluorescent dyes and quantum dots have been used extensively in this endeavor; however they each possess their own limitations in terms of photo-stability and/or cytotoxicity, two key parameters for long term *in vivo* studies. Over the past decade, there has been increased interest in the application of a new fluorescent probe in the form of nanodiamonds [7-9]. Atomic defects present in diamond have been shown to be photo-stable [7], bright and suitable for long term SPT applications [10, 11].

The long term stability and demonstrated bio-compatibility of nanodiamonds [12, 13] has opened up new opportunities for *in vivo* imaging [14, 15]. Nanodiamonds with a particular atomic defect namely, the negatively charged nitrogen-vacancy (NV) centre can be individually identified and bar coded via their optically detected magnetic resonance spectrum [11], which could see them applied to long term cell lineage studies. However, before these types of applications can be fully explored we need to explore the applicability of these probes in model developmental biological systems. Here we take an important step in this endeavor by incorporating and imaging individual fluorescent nanodiamonds during embryogenesis in the genetic model organism, *Drosophila melanogaster*. We demonstrate a technique for incorporating nanodiamonds in the blastoderm cells of the developing embryo and implement single particle tracking (SPT) techniques to characterise the motion and forces applied on the nanodiamonds in discrete areas of the embryo up to a depth of 40 µm.

The development of the *Drosophila* embryo begins with 13 rounds of mitosis in a single syncytium, which results in a single layer of nuclei underlying the plasma membrane. Cellularisation begins with furrows of plasma membrane introgressing between adjacent nuclei and enclosing each nucleus to form a layer of outer cells (blastoderm cells) and a single large syncytial yolk cell as shown in Figure 1. The accessibility of these outer cells for live imaging, combined with the ability to manipulate cellular processes both genetically and pharmacologically (i.e. by injection of drugs into the syncytium) [16-18] has made cellularisation an outstanding model system for studying the complex cellular processes underlying embryonic development [19, 20].

In this work, we take advantage of the photo-stable properties of the NV defect in diamond and demonstrate SPT of individual nanodiamonds in the blastoderm cells of developing *Drosophila* embryos. Fluorescence correlation spectroscopy (FCS) and wide-field microscopy are employed to study the dynamics of individual nanodiamonds at a range of depths throughout stage 5 of development. These techniques allow properties such as the diffusion, average velocity and applied force on the nanoparticles to be determined.

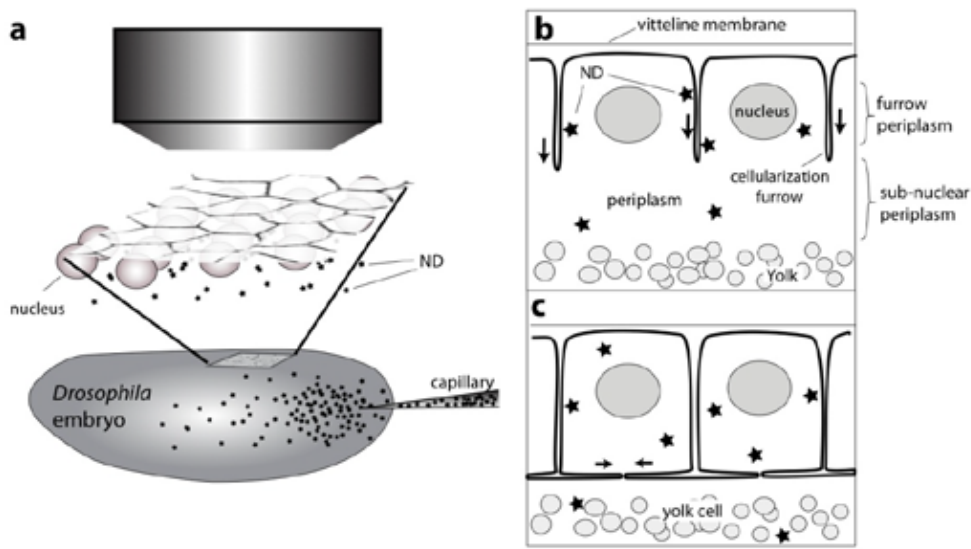

Fig. 1: a. Schematic of the micro-injection of nanodiamonds into the *Drosophila* embryo. b-c. Early (b) and late (c) stage 5 embryos showing the cellularisation furrows introgressing between nuclei, which invade the yolk-free periplasm during the later syncytial divisions (b, arrows). The ingressing membranes eventually join and pinch off individual blastoderm cells, forming as a consequence a large, internal yolk cell. Nanodiamonds that have diffused into the yolk-free periplasm can become internalised in the blastoderm cells at the completion of stage 5 (c).

## 2. Experimental Section

*Drosophila* adults were placed in an egg-laying chamber at 25°C and embryos collected after 55 minutes. Embryos were bleached for 3 minutes (50% household bleach) to remove the waxy, opaque, outer covering (chorion), rinsed in tepid water, mounted on a coverslip using a thin layer of rubber cement and allowed to desiccate for 3-4 mins before being covered with Halocarbon oil. Micro-injection of the nanodiamond suspension was performed with Clark GC100TF-10 capillaries pulled on a Brown-Flaming Sutter P80 puller, or FemtoTip II pre-made tips (Eppendorf), and a FemtoJet Express pressure injector (Eppendorf). A volume of liquid 50-75 µl was injected into the posterior third of a pre-cellularisation embryo and allowed to diffuse throughout the syncytial cytoplasm. Embryos were allowed to develop for 60-90 mins before being imaged. A schematic representation of the experimental arrangement is shown in Figure 1.

The synthesis of bright fluorescent nanodiamonds was conducted from commercially available Type IIa material (SYP 0.1) from Van Moppes, Switzerland. The source material has an initially high nitrogen concentration ([N] ~ 4.8 ppm) determined from electronic paramagnetic resonance studies. In order to improve the brightness of the diamonds the source material was irradiated with 2 MeV electrons with a fluence of $1\times10^{18}$cm$^{-2}$ at a temperature of < 80°C in a nitrogen ambient atmosphere. The induced vacancies were then made mobile to promote the formation of additional NV centres in the diamond lattice. This was achieved by a high temperature anneal of the irradiated material at 800°C for 2 hours in a vacuum of $10^{-7}$ Torr. The nanodiamond powder was then oxidized at 425°C for 2 hours dispersed in milliQ water (1mg/ml), sonicated for 36 hours and centrifuged at 12,000 rcf for 2 min. The supernatant solution was examined using the Zetasizer nano (Malvern) and was found to exhibit an average particle distribution size of 131 ± 60 nm with a zeta potential of –28.5 mV representing a stable colloidal solution. Before micro-injection the nanodiamonds we conjugated with bovine serum albumin (BSA) to reduce bio-fouling at a ratio of 1:1 in milliQ water at a concentration of 0.5% w/v. Particle sizing and zeta potential measurements after conjugation revealed no significant change in particle size but an increase in the zeta potential to -39.7 mV.

Imaging was performed on an inverted confocal fluorescence microscope (Nikon, Eclipse Ti-U) with a ´100 1.45 NA (Nikon) oil immersion objective operating with 532nm excitation. Fluorescence from the nanodiamonds was collected with a ×100 1.45 NA objective and filtered using a long pass (560 nm) and band pass (650-750 nm) filter (Semrock) to eliminate unwanted pump excitation. The fluorescence was then focused onto a multi-mode optical fibre (core 62.5 μm) which acts as a pinhole for the confocal microscope. The sampling volume of the confocal microscope was determined through the measurement of the lateral and spatial resolution of the microscope. This was done by dispersing 15 nm nanodiamonds onto a glass coverslip. Nanodiamonds containing single NV centres act as point sources of light. Line scans in *x,y* and *z* were performed on 20 different single NV centres and fitted to a Gaussian profile. The full width half maximum of the averaged line scans was taken as the lateral and axial resolution. The lateral and spatial resolution of the microscope was $r_0$ = 290 ± 40 nm and $z_0$ = 490 ± 85 nm, respectively, which compared well with the manufactures specifications for the oil immersion lens.

*Drosophila melanogaster* embryos were imaged in two optical configurations: confocal and wide-field microscopy. Confocal imaging was performed with 532 nm excitation (300 μW) in a temperature controlled environment of 18ºC. The fluorescence from individual nanodiamonds was detected using a single photon counting detector (Perkin Elmer, SPCM-AQRH-12) with photon arrival times correlated with a multiple event time digitizer (FAST ComTec, P7889). Wide-field imaging was performed on the same inverted microscope with the addition of a telescope to expand the excitation beam by a factor of 3 and a focusing lens (f = 300 mm) to focus the excitation light onto the back aperture of the objective creating a uniform wide-field illumination. Typical excitation powers used for wide-field imaging ranged between 2-4 W/mm$^2$. The wide-field fluorescence image was detected with a sCMOS camera (Andor, Neo) over the 60×60 μm field of view.

### 3. Results and discussion

*In vivo* imaging of the embryos was undertaken during stage 5 (cellularisation) of embryonic development. The fluorescent nanodiamonds were clearly observed above any cell auto fluorescence with a signal to noise ratio (SNR) of 100 (10) for confocal (wide-field) imaging. This strong signal allowed convenient particle tracking techniques, FCS and wide-field SPT, to be employed to probe the subcellular dynamics. In the first part of the investigation, FCS was used to characterize the dynamics of individual fluorescent nanodiamonds in the blastoderm cells during stage 5 of development. At this point the cells are not completely isolated and still maintain connections with the syncytial yolk cytoplasm through wide cytoplasmic bridges. A typical confocal image of injected nanodiamonds in the blastoderm cells at the posterior end of the embryo 10 μm above the cover slip is shown in Figure 2.

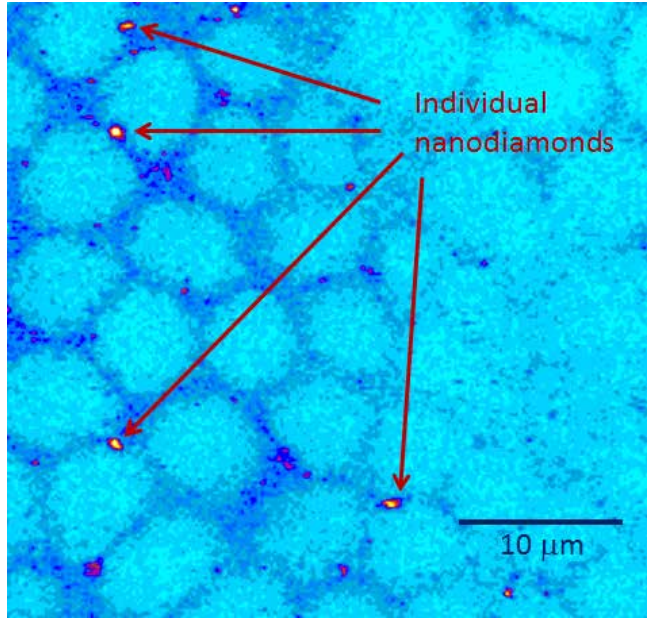

Fig. 2. Scanning confocal fluorescence image of individual nanodiamonds in the blastoderm cells during stage 5 of development. Note: The majority of nanodiamonds observed are localized to the cell periphery.

The cell membrane auto-fluorescence shows the arrangement of the blastoderm cells in a honeycomb type structure. The dwell time per pixel was set to 10 ms to maximize the signal from the fluorescent nanodiamonds. This led to non-Gaussian fluorescence images of the individual nanodiamonds due to diffusion over the 2 sec intervals between line scans. Fluorescent nanodiamonds located within this field of view were targeted for FCS studies. As shown in Figure 2 the majority of nanodiamonds are localised to the periphery of the cell where the membrane furrow is introgressing. With the confocal excitation beam fixed on a single pixel, fluorescence counts were monitored and correlated over time and in general could be described by the auto-correlation function for free diffusion in two dimensions [21]:

$$G(\tau) = \frac{1}{\bar{N}}\left(1 + \frac{\tau}{\tau_d}\right)^{-1} \cdot \left(1 + \frac{\tau}{\omega^2 \tau_d}\right)^{-1/2} \qquad (1)$$

where $t_d$ is the average residual time in the focal volume under free diffusion, $\omega = z_0 / r_0$ is the aspect ratio of the sampling volume and $\bar{N}$ is the average number of molecules in the sampling volume. The diffusion co-efficient, $D$, of a particle is then given by the relationship $D = r_0^2/4t_d$.

A typical fluorescence intensity correlation plot obtained from an individual fluorescent nanodiamond in a blastoderm cell is shown in Figure 3. The fluorescence signal from the nanodiamond was filtered using a long pass (560 nm) and band-pass filter (650-750 nm) and measured with the single photon counting detector and multiple event time digitizer. The fluorescence counts were recorded as a function of time using the event time digitizer with a bin size of 13749 ns.

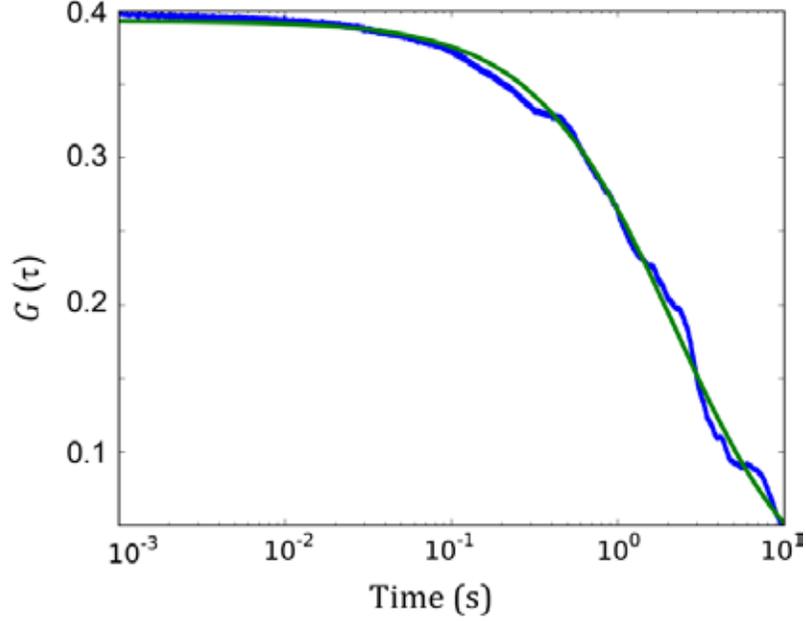

Fig. 3. Typical FCS from a freely diffusing nanodiamond in a blastoderm cell (solid blue line). The solid green line is a fit to the data using Equation (1) with the exacted parameter $D = (8.27 \pm 0.01) \times 10^{-3}$ µm²/s.

The concentration of fluorescent nanodiamonds was kept sufficiently low such that on average only one nanodiamond was diffusing through the excited volume at once as confirmed by the correlation amplitude of ~1. The mean diffusion coefficient extracted from 15 motility events was $D = (6 \pm 4) \times 10^{-3}$ µm²/s, (mean ± SD). Of the nanodiamonds probed using FCS more than 80% were found to undergo free diffusion. Driven motion of individual nanodiamonds was also observed with an example shown in Figure 4. For driven motion the intensity correlation data were modeled by:

$$G(\tau) = \frac{1}{N}\left(1 + \frac{\tau}{\tau_d}\right)^{-1} \cdot \left(1 + \frac{\tau}{\omega^2 \tau_d}\right)^{-1/2} \exp\left[\left(-\frac{\tau}{\tau_v}\right)^2 \cdot \frac{1}{1+\frac{\tau}{\tau_d}}\right] \qquad (2)$$

where $t_v = r_0/v$ is the average residual time if there is only a flow (no diffusion) and $v$ is the velocity of the particle.

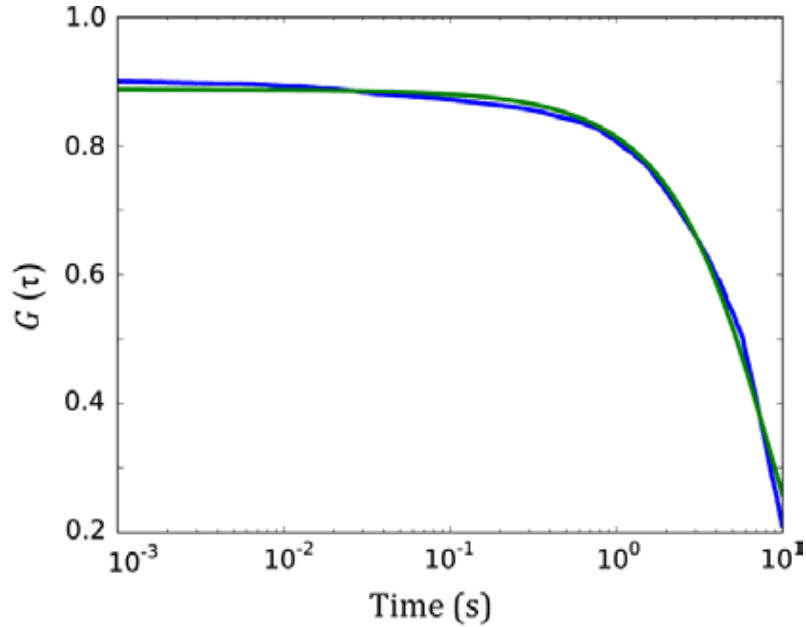

Fig. 4: FCS from a nanodiamond undergoing driven motion (solid blue line). The solid green line is a fit to the data using Equation 2. The extracted diffusion co-efficient and velocity from the fit was $D = (1.41 \pm 0.01) \times 10^{-3}$ µm$^2$/s and $v = 0.030 \pm 0.003$ µm/s.

The diffusion co-efficient for the nanodiamond shown in Figure 4 was found to be $D = 1.41 \pm 0.01 \times 10^{-3}$ µm$^2$/s with a velocity of $v = r_0/t_v = 0.030 \pm 0.003$ µm/s. With the knowledge of the driven velocity of the nanodiamond, information regarding the applied force on the nanoparticle can also be determined using a combination of the Stokes-Einstein relationship $D = k_b T/6\pi\eta R$ and Stokes Law $f_e = 6\pi\eta R v$:

$$f_e = v \cdot k_b \cdot T / D \qquad (3)$$

where $v$ is the velocity of the particle, $k_b$ is the Boltzmann constant, $T$ is the temperature, $D$ is the diffusion co-efficient R is the is the dynamic viscosity and $\eta$ is the radius of the particle.

Therefore, the applied force on the nanodiamond was $0.08 \pm 0.01$ pN. Driven motion in this model system can be attributed to cytoskeletal motors such as dynein, kinesin and myosin which have been shown to generate forces *in vitro* from 0.25 to 5 pN, depending on the applied load and local ATP concentration [1, 22]. The significantly smaller forces observed here *in vivo* may be a result of the dense cytoskeletal networks present in developing blastoderm cells.

*3.1 Wide-field imaging of nanodiamonds in vivo*

Although FCS provides an accurate and detailed description of the intra-cellular dynamics of nanoparticles, its small sampling volume, in our case $0.23 \pm 0.08$ µm$^3$, is limiting for probing the large number of particles over wide fields of view. An alternative technique to FCS is single particle tracking through wide-field microscopy, whereby sequential images of the entire field of view can be acquired and the trajectory of many nanoparticles can be tracked simultaneously over long periods of time. By tracking the individual trajectories of nanodiamonds and determining the mean squared displacement over time we can again probe the intra-cellular dynamics of the individual nanodiamonds. Figure 5 shows a wide-field fluorescence image of a stage 5 *Drosophila* embryo injected with fluorescent nanodiamonds. The nanodiamonds are clearly observed above the embryo auto-fluorescence, allowing us to efficiently probe the dynamics of individual nanodiamonds in distinct regions of the embryo. Figure 5 also highlights the trajectories of the nanodiamonds over a time period of 5s. The trajectories were identified from sequential images with a 200 ms acquisition time using ImarisCell (Bitplane) spot tracker software package. Particle trajectories from individual nanodiamonds where determined from particles which appeared consistently in 5 consecutive images and were isolated from other surrounding particles, to rule out any particle identity confusion.

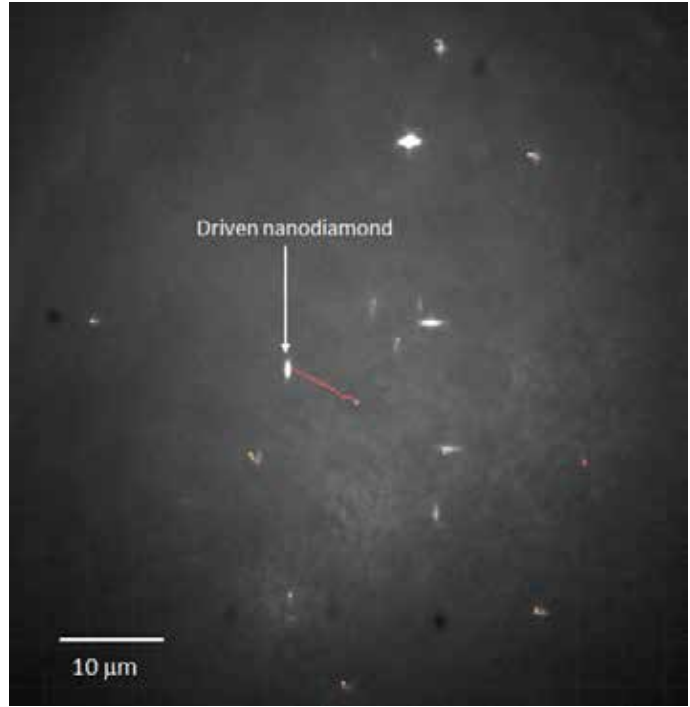

Fig. 5. Wide-field fluorescence image from nanodiamonds micro-injected into a *Drosophila melanogaster* embryo. Individual nanodiamonds are identified by the circles in the image. The red lines represent the trajectories of the nanodiamonds over a 5 second period.

The mean squared displacement (MSD) of each nanodiamond can then be determined from the trajectory data. In particular, the slope of the MSD versus time is directly related to the diffusion coefficient and driven motion for that particle. This relationship is found from the Einstein, Fokker-Planck or Langevin theory of Brownian motion [26] and is given by:

$$\langle |x|^2 \rangle = \sum_{k=1}^{N} |x_k|^2 = (2n)Dt + (v_x t)^2 \qquad (4)$$

where *n* represents the displacement in 1, 2 or 3 dimensions and $v_x$ is the drift velocity in the *x* direction arising from an external force providing a driven unidirectional component to the random motion of a particle.

Wide-field fluorescence images were acquired of the lateral surface of the embryo towards the posterior end, at several depths within the embryo. The first sets of images were measured at the level of the nuclei and ingressing furrow (10-15 µm from the outer membrane) (i.e. hereafter "furrow periplasm") and a second set of images in the periplasm underlying the nuclei (20-40 µm from the outer membrane, hereafter the "sub-nuclear periplasm"). Typical nanodiamond trajectories and MSDs over time are shown in Figure 6 for nanodiamonds within the furrow periplasm during stage 5.

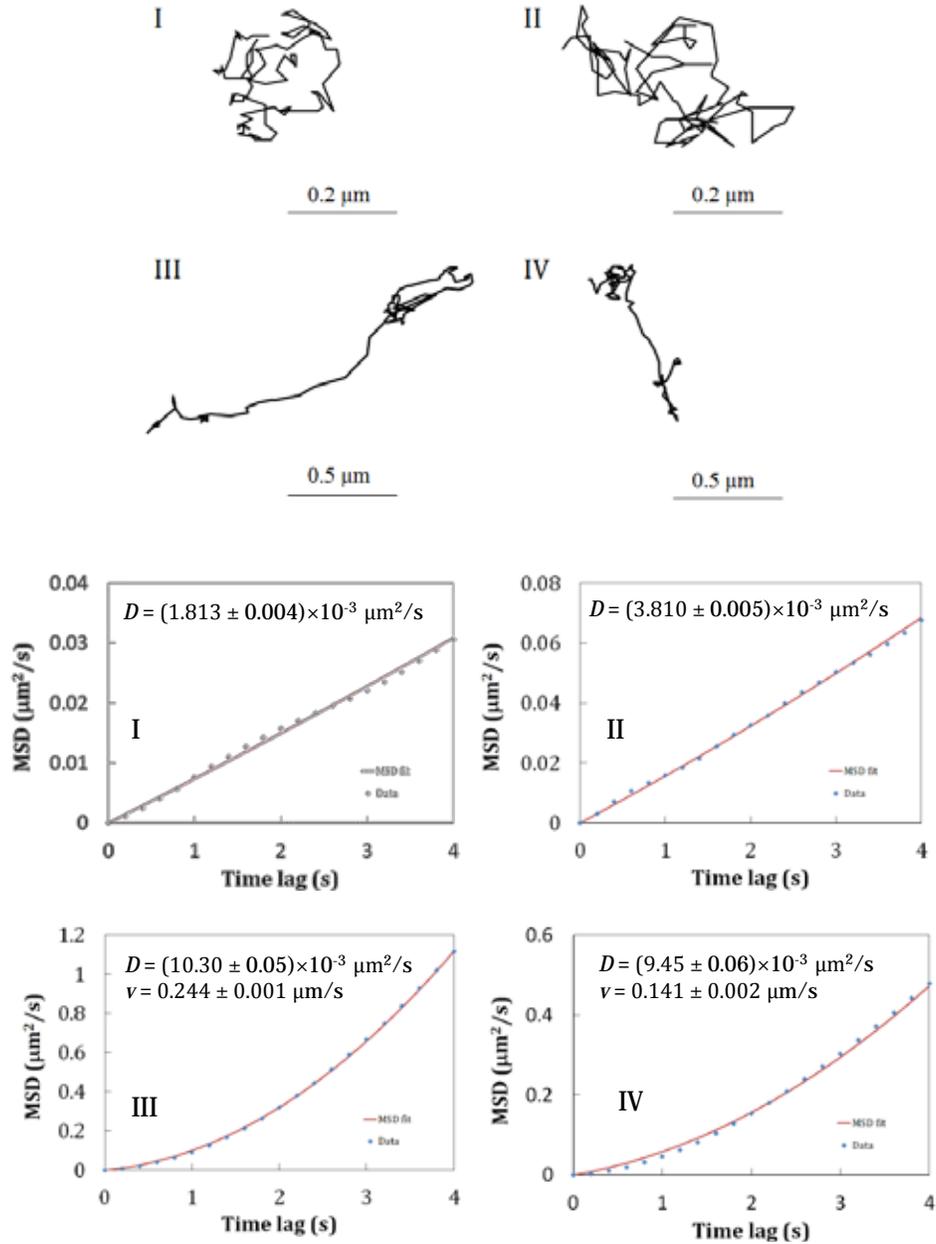

Fig. 6: Particle trajectories from four individual nanodiamonds (I-IV) in the furrow periplasm during stage 5 of development. The mean squared displacement as a function of time is shown below for each respective trajectory (I-IV).

Sixty percent of fluorescent nanodiamonds observed within the furrow periplasm at the posterior end of the embryo were found to undergo free diffusion with linear MSDs and a mean diffusion co-efficient of $(6 \pm 3) \times 10^{-3}$ μm$^2$/s (mean ± SD) from 23 motility events. The broad standard deviation in our case arises from the non-uniform particle size distribution 130 ± 60 nm. 40% of the nanodiamonds were found to undergo driven motion as evidenced by their non-linear MSDs. By fitting the data to Equation 4 the diffusion co-efficient, driven velocity and applied force was determined. The mean driven velocity observed was $v_x = 0.13 \pm 0.10$ μm/s (mean ± SD) from 19 motility events. Taking into account the measured diffusion co-efficient of each particle, the mean applied force exerted on an individual nanodiamond was 0.07 ± 0.05 pN (mean ± SD) with a maximum applied force of 0.21 pN. These applied forces are again consistent with the forces generated by cytoskeletal motors. The measured driven velocities of the nanodiamonds are also consistent with the measured velocities of kinesin, dynein and myosin motors which have been shown to vary from 0.1 to 1 um/s for loads from 1 to 5 pN [1, 22, 23]. One particular example highlighted in

Figure 5 shows a single nanodiamond being trafficked at much faster speeds $v_x = 0.50 \pm 0.02$ µm/s. This demonstrates large differences in the local forces and viscosities within the developing blastoderm cells throughout stage 5.

As a comparison, the dynamics of the injected particles in the sub-nuclear periplasm were also observed. The majority of nanodiamonds in the sub-nuclear periplasm (60%) were found to undergo some form of driven motion with 40% of the nanodiamond exhibiting free diffusion. The mean diffusion co-efficient of the nanodiamonds in the sub-nuclear periplasm was $(63 \pm 35) \times 10^{-3}$ µm$^2$/s (mean ± SD) from 13 motility events. This is on average a factor of 10 faster than the diffusion rates observed in the furrow periplasm. The difference in the diffusion coefficients in the furrow periplasm and sub-nuclear periplasm may be attributed to the fact that the majority of nanodiamonds observed in the furrow periplasm were located at the cell periphery and may be associated with dense cytoskeletal networks which could inhibit the motion.

The average driven velocity observed from nanodiamonds in the sub-nuclear periplasm was $0.27 \pm 0.12$ µm/s (mean ± SD) from 18 motility events which indicates an average applied force of $0.05 \pm 0.04$ pN (mean ± SD) with a maximum applied force of $0.28 \pm 0.02$ pN. This result shows that the transport mechanisms present within the blastoderm cells at both the furrow and sub-nuclear levels are analogous. Furthermore, this simple technique for determining the applied force on the nanodiamond may prove to be an effective tool for improving our understanding of embryonic trafficking and development.

*3.2 Embryo Injections and nanodiamond toxicity test*

Survivability tests were also carried out on a set of *Drosophila* embryos injected with separate milliQ, BSA and the conjugated nanodiamond:BSA suspension. *Drosophila melanogaster* embryos were allowed to develop for 1 day at 25˚C in a humidified chamber and the number of injected embryos that had hatched into the oil assayed. Survival rates for the nanodiamond in BSA solution (12.16%, n=148) did not differ significantly from BSA controls 14.07% (n=135) (p=0.7248, Fisher's Exact test). This is in agreement with previous toxicity studies carried out on nanodiamonds in *C. Elegans* embryos [12]. There was also no observable developmental delay in any group compared with un-injected embryos. This result further validates the applicability of florescent nanodiamond probes for long term intra-cellular studies *in vivo*.

**4. Conclusion**

In conclusion we have demonstrated that micro-injected nanodiamonds are effective nanoscale fluorescent probes for determining the intra-cellular dynamics in developing *Drosophila* embryos. FCS and wide-field imaging techniques were used to identify and track individual nanodiamonds. Free diffusion and driven motion of individual nanodiamonds were observed in blastoderm cells at the posterior end of *Drosophila* embryos during stage 5 of development. From the trajectories analysis we determined a mean diffusion co-efficient and driven velocity of $(6 \pm 3) \times 10^{-3}$ µm$^2$/s, (mean ± SD) and $0.13 \pm 0.10$ µm/s (mean ± SD) µm/s, respectively in the furrow periplasm. While in the sub-nuclear periplasm the mean diffusion co-efficient and driven velocity was observed $(63 \pm 35) \times 10^{-3}$ µm$^2$/s (mean ± SD) and $0.27 \pm 0.12$ µm/s, (mean ± SD) respectively. The mean applied force on the nanodiamonds ranged between 0.05 and 0.07 pN in the two regions of interest with a maximum applied force of $0.28 \pm 0.02$ pN, which suggests the transport mechanisms of the nanodiamonds in the furrow and sub-nuclear periplasm are analogous.

Future work to conjugate biological molecules to the nanodiamonds in order to visualise the dynamic behaviour of more targeted molecules such as kinesin motors is currently underway. This combined with the attractive quantum properties offered by nanodiamond probes may allow for more advanced tagging and orientation tracking [11, 24] applications, with the potential for cell lineage tracking as the embryo develops into more complicated stages.

**Acknowledgements**

The authors would like to acknowledge the Melbourne Materials Institute at the University of Melbourne for the seed funding to undertake this research. This research was supported in part by the Australian Research Council Centre of Excellence for Quantum Computation and Communication Technology (Project number CE110001027). MM and RS acknowledge support from the University of Melbourne and Australian Research Council Discovery grant (DP120104443). FC was supported by the Australian Research Council under the Australian Laureate

Fellowship (FL120100030) and YY by the Discovery Early Career Researcher Award (DE130100488). The authors acknowledge Professor Jörg Wrachtrup for helpful discussions and Philipp Senn for assistance with sonicating the nanodiamond suspensions.

**References**
1. K. Visscher, M. J. Schnitzer, and S. M. Block, "Single kinesin molecules studied with a molecular force clamp," Nature 400, 184-189 (1999).
2. M. P. Sheetz, and J. A. Spudich, "Movement of myosin-coated fluorescent beads on actin cables in vitro," Nature 303, 31-35 (1983).
3. C. Kural, H. Kim, S. Syed, G. Goshima, V. I. Gelfand, and P. R. Selvin, "Kinesin and Dynein Move a Peroxisome in Vivo: A Tug-of-War or Coordinated Movement?," Science 308, 1469-1472 (2005).
4. J. Eid, A. Fehr, J. Gray, K. Luong, J. Lyle, G. Otto, P. Peluso, D. Rank, P. Baybayan, B. Bettman, A. Bibillo, K. Bjornson, B. Chaudhuri, F. Christians, R. Cicero, S. Clark, R. Dalal, A. deWinter, J. Dixon, M. Foquet, A. Gaertner, P. Hardenbol, C. Heiner, K. Hester, D. Holden, G. Kearns, X. Kong, R. Kuse, Y. Lacroix, S. Lin, P. Lundquist, C. Ma, P. Marks, M. Maxham, D. Murphy, I. Park, T. Pham, M. Phillips, J. Roy, R. Sebra, G. Shen, J. Sorenson, A. Tomaney, K. Travers, M. Trulson, J. Vieceli, J. Wegener, D. Wu, A. Yang, D. Zaccarin, P. Zhao, F. Zhong, J. Korlach, and S. Turner, "Real-Time DNA Sequencing from Single Polymerase Molecules," Science 323, 133-138 (2009).
5. X. Michalet, F. Pinaud, L. Bentolila, J. Tsay, S. Doose, J. Li, G. Sundaresan, A. Wu, S. Gambhir, and S. Weiss, "Quantum dots for live cells, in vivo imaging, and diagnostics," Science 307, 538-544 (2005).
6. M. J. Saxton, and K. Jacobson, "Singe-particle tracking:Applications to Membrane Dynamics," Annual Review of Biophysics and Biomolecular Structure 26, 373-399 (1997).
7. C. C. Fu, H. Y. Lee, K. Chen, T. S. Lim, H. Y. Wu, P. K. Lin, P. K. Wei, P. H. Tsao, H. C. Chang, and W. Fann, "Characterization and application of single fluorescent nanodiamonds as cellular biomarkers," Proc. Natl Acad. Sci. USA 104, 727 (2007).
8. O. Faklaris, D. Garrot, V. Joshi, F. Druon, J. P. Boudou, T. Sauvage, P. Georges, P. A. Curmi, and F. Treussart, "Detection of single photoluminescent diamond nanoparticles in cells and study of the internalization pathway," Small 4, 2236 (2008).
9. O. Faklaris, V. Joshi, T. Irinopoulou, P. Tauc, M. Sennour, H. Girard, C. Gesset, J. C. Arnault, A. Thorel, J. P. Boudou, P. A. Curmi, and F. Treussart, "Photoluminescent diamond nanoparticles for cell labeling: study of the uptake mechanism in mammalian cells," ACS Nano 3, 3955 (2009).
10. Y. R. Chang, H. Y. Lee, K. Chen, C. C. Chang, D. S. Tsai, C. C. Fu, T. S. Lim, Y. K. Tzeng, C. Y. Fang, C. C. Han, H. C. Chang, and W. Fann, "Mass production and dynamic imaging of fluorescent nanodiamonds," Nat Nano 3, 284-288 (2008).
11. L. P. McGuinness, Y. Yan, A. Stacey, D. A. Simpson, L. T. Hall, D. Maclaurin, S. Prawer, P. Mulvaney, J. Wrachtrup, F. Caruso, R. E. Scholten, and L. C. L. Hollenberg, "Quantum measurement and orientation tracking of fluorescent nanodiamonds inside living cells," Nat. Nanotech. 6, 358-363 (2011).
12. N. Mohan, C.-S. Chen, H.-H. Hsieh, Y.-C. Wu, and H.-C. Chang, "In Vivo Imaging and Toxicity Assessments of Fluorescent Nanodiamonds in Caenorhabditis elegans," Nano Lett. 10, 3692-3699 (2010).
13. V. Vaijayanthimala, P.-Y. Cheng, S.-H. Yeh, K.-K. Liu, C.-H. Hsiao, J.-I. Chao, and H.-C. Chang, "The long-term stability and biocompatibility of fluorescent nanodiamond as an in vivo contrast agent," Biomaterials 33, 7794-7802 (2012).
14. Y. Kuo, T.-Y. Hsu, Y.-C. Wu, and H.-C. Chang, "Fluorescent nanodiamond as a probe for the intercellular transport of proteins in vivo," Biomaterials 34, 8352-8360 (2013).
15. R. Igarashi, Y. Yoshinari, H. Yokota, T. Sugi, F. Sugihara, K. Ikeda, H. Sumiya, S. Tsuji, I. Mori, H. Tochio, Y. Harada, and M. Shirakawa, "Real-Time Background-Free Selective Imaging of Fluorescent Nanodiamonds in Vivo," Nano Lett. 12, 5726-5732 (2012).
16. J. M. Crawford, N. Harden, T. Leung, L. Lim, and D. P. Kiehart, "Cellularization in Drosophila melanogaster Is Disrupted by the Inhibition of Rho Activity and the Activation of Cdc42 Function," Developmental biology 204, 151-164 (1998).
17. A. M. Sokac, and E. Wieschaus, "Local actin-dependent endocytosis is zygotically controlled to initiate Drosophila cellularization," Dev Cell 14, 775-786 (2008).
18. A. Royou, C. Field, J. C. Sisson, W. Sullivan, and R. Karess, "Reassessing the role and dynamics of nonmuscle myosin II during furrow formation in early Drosophila embryos," Mol Biol Cell 15, 838-850 (2004).

19. T. Lecuit, "Junctions and vesicular trafficking during Drosophila cellularization," J Cell Sci 117, 3427-3433 (2004).
20. A. Mazumdar, and M. Mazumdar, "How one becomes many: blastoderm cellularization in Drosophila melanogaster," Bioessays 24, 1012-1022 (2002).
21. O. Krichevsky, and G. Bonnet, "Fluorescence correlation spectroscopy: the technique and its applications," Rep. Prog. Phys. 65, 251 (2002).
22. R. Mallik, B. C. Carter, S. A. Lex, S. J. King, and S. P. Gross, "Cytoplasmic dynein functions as a gear in response to load," Nature 427, 649-652 (2004).
23. M. P. Sheetz, and J. A. Spudich, "Movement of myosin-coated fluorescent beads on actin cables in vitro," (1983).
24. E. H. Chen, O. Gaathon, M. E. Trusheim, and D. Englund, "Wide-Field Multispectral Super-Resolution Imaging Using Spin-Dependent Fluorescence in Nanodiamonds," Nano Lett. 13, 2073-2077 (2013).